\documentclass{Interspeech}

\interspeechcameraready

\title{
Teaching Audio-Aware Large Language Models What Does Not Hear: Mitigating Hallucinations through Synthesized Negative Samples
}

\author[]{Chun-Yi}{Kuan}
\author[]{Hung-yi}{Lee}

\affiliation[nocounter]{Graduate Institute of Communication Engineering}{National Taiwan University}{Taiwan}
\email{chunyi.kuan.tw@gmail.com, hungyilee@ntu.edu.tw}

\usepackage{comment}
\usepackage{cite}
\usepackage{hyperref}
\usepackage{url} 
\usepackage{amsmath,amssymb,amsfonts}
\usepackage{algorithmic}
\usepackage{graphicx}
\usepackage{textcomp}
\usepackage{booktabs}
\usepackage{soul}
\usepackage{xcolor}
\usepackage{xurl}
\usepackage{bm}

\definecolor{YInMnBlue}{HTML}{4E598C}
\definecolor{LapisBlue}{HTML}{255F85}
\definecolor{CelestialBlue}{HTML}{009DDC}
\definecolor{IndigoBlue}{HTML}{344966}
\definecolor{ResolutionBlue}{HTML}{14248A}

\begin{document}

\maketitle

\begin{abstract}
Recent advancements in audio-aware large language models (ALLMs) enable them to process and understand audio inputs. 
However, these models often hallucinate non-existent sound events, reducing their reliability in real-world applications. 
To address this, we propose LISTEN (Learning to Identify Sounds Through Extended Negative Samples), a contrastive-like training method that enhances ALLMs’ ability to distinguish between present and absent sounds using synthesized data from the backbone LLM. 
Unlike prior approaches, our method requires no modification to LLM parameters and efficiently integrates audio representations via a lightweight adapter. 
Experiments show that LISTEN effectively mitigates hallucinations while maintaining impressive performance on existing audio question and reasoning benchmarks. 
At the same time, it is more efficient in both data and computation.

\end{abstract}

\section{Introduction}

Audio-aware large language models (ALLMs) expand on traditional text-based LLMs by incorporating the ability to process audio. 
These models can process audio, speech, and text inputs at the same time, using text prompts to extract relevant information from audio and speech. 
This advancement enables large language models to understand multimodal information and interact with users by extracting information based on user instructions. 
However, recent studies \cite{kuan2024can, kuan2024understanding} have found that ALLMs, like text-based LLMs~\cite{rawte2023survey,zhang2023siren,huang2023survey,friel2023chainpoll,liu2024survey}, are prone to hallucination issues. 
For example, they may incorrectly identify sound events that do not exist in the audio, affecting their reliability and safety in understanding and processing audio. 
A model might, for instance, mistake a siren for music, leading to missed alerts in an emergency response system and potentially delaying critical actions.

To address the hallucination issues in models, we propose a contrastive-like training method that not only teaches the model to recognize sounds present in the audio but also helps it learn which sounds are absent by leveraging data synthesized by the backbone LLMs.
Unlike traditional methods for training ALLMs, which often require extensive audio instruction-following datasets to align audio and text modalities, our approach uses a simple, efficient, and effective automatic pipeline inspired by previous works\cite{fathullah2023towards, lu2024developing, wang2023blsp, wang2024blsp} to generate audio-text pairs. 
Additionally, this pipeline also produces contrastive data, incorporating both positive and negative training samples to inject audio understanding into the model.

Our experimental results show that our method not only mitigates audio hallucinations but also achieves competitive performance on audio question-answering benchmarks and advanced audio reasoning tasks. We find that using self-generated data is beneficial, and incorporating negative samples within this synthesized data further improves the model’s ability to reduce audio hallucinations.
In addition, the total duration of training data required is only 3\% of the largest dataset used by the baseline model with the most extensive training data. 

Compared to the most common approaches\cite{chu2023qwen, chu2024qwen2, tang2023salmonn, gong2023joint} that incorporate LoRA\cite{hulora} into LLMs for training or even fine-tune the entire model, our method does not add or modify any parameters of the large language model during training.
Instead, inspired by the previous work\cite{zhuminigpt, lu2024desta, wang2023blsp}, we train a simple adapter that maps audio representations to the input dimensions of the LLM. 
This approach makes our method more efficient in terms of both training data and training time. 
We provide illustrative examples on our demo page~\footnote{\url{https://kuan2jiu99.github.io/Balsa}}.
In conclusion, our contributions are outlined as follows:
\begin{enumerate}
    \item \textbf{Hallucination Mitigation:} 
    This work is the first to address hallucination in audio-aware large language models.
    To tackle this issue, we propose \textbf{LISTEN} (\textbf{L}earning to \textbf{I}dentify \textbf{S}ounds \textbf{T}hrough \textbf{E}xtended \textbf{N}egative \textbf{S}amples), a contrastive-like method that helps models distinguish between present and absent sounds, effectively reducing hallucination and improving reliability.
    Our findings show that self-generated data plays a crucial role, and incorporating negative samples further enhances the model’s ability to mitigate hallucinations.

    \item \textbf{Efficient Data Utilization and Training:}  
Our method integrates general audio representations into LLMs using a lightweight adapter without modifying any LLM parameters, requiring only 3\% to 30\% of the dataset used by baseline models.  
This is achieved by leveraging a backbone-LLM-synthesized dataset, which automatically generates audio-text pairs and contrastive data across general audio scenarios.  
This process eliminates the need for extensive instruction-following dataset, ensuring robust learning while minimizing data requirements.

\end{enumerate}

\begin{figure*}[ht]
    \centering
    \includegraphics[width=1.0\textwidth]{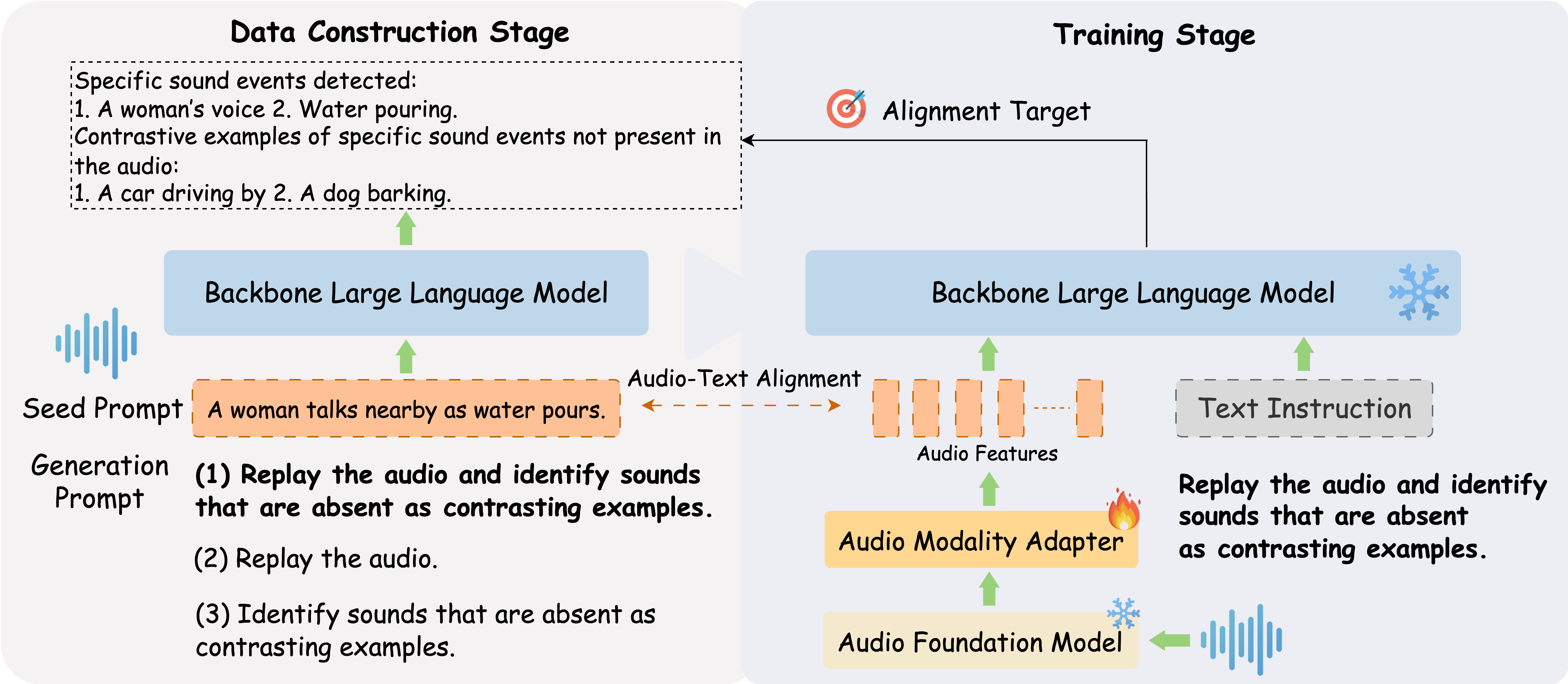} 
    \caption{
    Data Construction Stage uses a backbone large language model to generate audio-text aligned descriptions. 
    Training Stage focuses on audio-text alignment by training an audio modality adapter, while the backbone large language model remains frozen.
    }
    \vspace{-5pt}
    \label{fig:overview} 
\end{figure*}
\vspace{-5pt}
\section{Related Work}
Audio-aware large language models (ALLMs) extend traditional LLMs by incorporating audio perception, enabling them to process both audio and text inputs for a wider range of audio-related tasks. Recent advancements~\cite{chu2023qwen, chu2024qwen2, tang2023salmonn, gong2023joint, lu2024desta, lu2024developing, kuan2024speech, fathullah2023towards, wang2023blsp, huang2024dynamic, huang2024dynamic2} have focused on bridging modality gaps by generating paired audio-text data, often leveraging powerful large language models~\cite{achiam2023gpt} to create training datasets. For instance, these models generate question-answering format data based on the metadata in audio datasets.
While many approaches rely on external LLMs for data generation, our method takes a different approach by utilizing the backbone LLM of the ALLM itself, similar to previous work~\cite{lu2024desta, wang2023blsp, fathullah2023towards}.  
Although our approach follows a similar strategy, we are the first to apply it to general audio tasks, demonstrating its effectiveness. 
In addition, we explore the use of self-generated data and synthesized negative samples, further enhancing the model’s learning process. 
This eliminates text discrepancies, leverages the model’s inherent understanding of audio dataset metadata, and reduces the need for complex and extensive training datasets, significantly improving data efficiency.


Despite these advancements, hallucination remains a major challenge in ALLMs, where models misidentify non-existent sound events, limiting their reliability~\cite{kuan2024can, kuan2024understanding} in practical applications. 
While prior studies have noted this issue, none have addressed it from a model training perspective. 
Our work addresses this gap by improving ALLM reliability through a contrastive-like training approach to reduce hallucinations.

\section{Method}
\subsection{Data Construction}

During the data construction stage, inspired by\cite{fathullah2023towards, lu2024developing}, our goal is to create audio-text pair data with minimal textual discrepancies between the underlying LLM used in the ALLM and the constructed training data. 
While LLMs cannot directly process audio, they excel at understanding and interpreting textual metadata\cite{kuan2024understanding}. 
Therefore, we utilize the meta-information from the original audio datasets, which commonly include two types of annotations: human-annotated captions (${D}_{{caption}}$) and sound event tags (${D}_{{tag}}$).

To further minimize textual discrepancies and avoid the need to specifically design question-answer pairs for formatting the training data, we developed a structured format called the seed prompt (${P}_{seed}$). 
The seed prompt could be either audio captions or sound event tags, depending on the annotation format of the original dataset.
We then append the generation prompts (${P}_{gen}$) to the seed prompt and feed it into the LLM, enabling it to generate descriptions that match the audio. 
There are three types of generation prompts:
\begin{enumerate}
    \item \textbf{Positive Samples Generation Prompt (${P}_{{pos}}$)} aims to generate descriptions of sound events that actually occur in the audio. 
    For example, \textit{Replay the audio}.
        \item \textbf{Negative Samples Generation Prompt (${P}_{{neg}}$)} aims to generate descriptions of sound events that are not present in the audio.
    For example, \textit{Identify sounds that are absent as contrasting examples}.
        \item \textbf{Combined Samples Generation Prompt (${P}_{{comb}}$)} integrates both of the above prompts. 
    It aims to generate descriptions of both the sound events that are present and those that are absent in the audio.
    For example, \textit{Replay the audio and identify sounds that are absent as contrasting examples}.
\end{enumerate}

In summary, the final input prompt (${P}_{final}$) to the LLM can be expressed in Equation (\ref{equation-prompt}). 
\begin{equation}
\begin{split}
P_{final} = & \, [\textit{Begin of audio}] \, P_{seed} \, [\textit{End of audio}] \, P_{gen}, \\
& P_{seed} \in \{D_{caption}, D_{tag}\}, \\
& P_{gen} \in \{P_{pos}, P_{neg}, P_{comb}\}.
\end{split}
\label{equation-prompt}
\end{equation}
In Equation (\ref{equation-prompt}), \textit{[Begin of audio]} and \textit{[End of audio]} are expressed in natural language format, with the purpose of indicating to the model that the information between them represents the content from the audio. 
When training audio-aware large language models, the position where $P_{seed}$ is located will be replaced with audio representation.
These generated responses (${R}_{LLM}$) to the input context ${P}_{final}$ are then used as our training targets.
Actual examples are shown in Table\ref{tab:prompt}.



\begin{table}[t]
    \small
    \centering
    \caption{Examples generated under different sampling prompts.}
    \begin{tabular}{p{23em}}
    \toprule
    \textbf{Input Audio:} \\
    \textcolor{ResolutionBlue}{Audio Caption: A woman talks nearby as water pours.} \\
    \hline
    \textbf{Positive Samples Generation Prompt ($\boldsymbol{P}_{\boldsymbol{pos}}$):} \\
    Replay the audio. \\
    \textbf{Response:} \\
    Water pouring sounds, woman talking in the background \\
    \hline
    \textbf{Negative Samples Generation Prompt ($\boldsymbol{P}_{\boldsymbol{neg}}$):} \\
    Identify sounds that are absent as contrasting examples. \\
    \textbf{Response:} \\
    Based on the provided audio, here are some specific sound events that are not present in the audio: \\
    1. A car driving by \hspace{0.1cm} 2. Birds chirping \hspace{0.1cm} 3. A dog barking \\
    \hline
    \textbf{Combined Samples Generation Prompt ($\boldsymbol{P}_{\boldsymbol{comb}}$):} \\
    Replay the audio and identify sounds that are absent as contrasting examples. \\
    \textbf{Response:} \\
    Specific sound events detected in the provided audio: \\
    1. A woman's voice \hspace{0.1cm} 2. Water pouring \\
    Contrastive examples of specific sound events not present in the provided audio: \\
    1. A car driving by \hspace{0.1cm} 2. A dog barking \hspace{0.1cm} 3. The door bang \\
    \bottomrule
\end{tabular}
\vspace{-10pt}
\label{tab:prompt}
\end{table}

\subsection{Model Design and Training Approach}

Figure \ref{fig:overview} illustrates the architecture of our model.
\begin{enumerate}
    \item \textbf{Audio Encoder}. 
    We utilize the foundation model Whisper\footnote{\scriptsize{\url{huggingface.co/openai/whisper-small}}}\cite{radford2023robust} as our audio encoder.
    Prior studies\cite{gong_whisperat, ma2023investigating} have highlighted Whisper’s impressive performance on various audio-related tasks, despite its original design for automatic speech recognition and speech translation tasks. 
    As Whisper follows an encoder-decoder architecture, we utilize only its encoder component. 
    To retain the advantages of the pre-trained model, the audio encoder's parameters remain frozen.
    
    \item \textbf{Backbone Large Language Model.} 
    This study employs the instruction-tuned LLaMA-3.1-8B\footnote{\scriptsize{\url{huggingface.co/meta-llama/Llama-3.1-8B-Instruct}}}\cite{dubey2024llama} as the core large language model. 
    To preserve the model's original text-processing capabilities, we opted not to apply LoRA\cite{hulora} or fine-tune any of its parameters.
    
    \item \textbf{Audio Modality Adapter.}
    The only trainable component is the audio modality adapter, which is randomly initialized. 
    This adapter projects the output representations extracted by the audio encoder into the input dimension of the backbone large language model. 
    In detail, we employ a Qformer\cite{li2023blip} architecture to extract audio features from the intermediate layers of the Whisper encoder. 
    These features are aggregated through a weighted sum with learnable weights, followed by a simple linear projection layer that aligns the aggregated features with the input dimensions of the backbone large language model.

\end{enumerate}

During training, the continuous audio features produced by the audio modality adapter replace the seed prompt (${P}{seed}$) in the template (see Eq.(\ref{equation-prompt})). The entire architecture is then optimized end-to-end using a next-token prediction loss, with the generated responses (${R}_{LLM}$) as the learning target. This approach enables the large language model to achieve audio-text alignment while only fine-tuning the parameters of the audio modality adapter.

\vspace{-5pt}
\section{Experiment}

\subsection{Training Datasets}

The training datasets we used include AudioSet-20K\cite{7952261}, AudioCaps\cite{kim2019audiocaps}, FSD50K\cite{fonseca2021fsd50k}, MACS\cite{8682858}, ESC50\cite{piczak2015dataset}, UrbanSound8K\cite{Salamon:UrbanSound:ACMMM:14}, Clotho\cite{lipping2022clotho}, and VocalSound\cite{gong_vocalsound}. 
Among these original datasets, AudioCaps, Clotho, and MACS contain ground truth labels in the form of human-annotated audio captions, while the remaining datasets have ground truth labels in the form of sound events tags. 
In Table~\ref{tab:dataset_summary}, we present the statistical data of our training datasets and compare them with the amount of data used in previous work.
The duration is calculated based on each unique audio, excluding any duplicate segments. 
The number of samples represents the number of data instances used during training.

\begin{table}[ht]
\footnotesize
\centering
\caption{Summary of training datasets.}
\begin{tabular}{lccc}
\hline
\textbf{Dataset}          
& \textbf{Duration (Hours)} & \textbf{\# Samples} \\ 
\hline
AudioSet-20K\cite{7952261} & 60.4 & 42K \\
AudioCaps\cite{kim2019audiocaps} & 138.0 & 96K \\
Clotho\cite{lipping2022clotho} & 18.0 & 27K \\
MACS\cite{8682858} & 10.9 & 33K \\
FSD50K\cite{fonseca2021fsd50k} & 56.4 & 56K \\
ESC50\cite{piczak2015dataset} & 2.8 & 4K \\
UrbanSound8K\cite{Salamon:UrbanSound:ACMMM:14} & 9.7 & 17K \\
VocalSound\cite{gong_vocalsound} & 20.2 & 34K \\
\hline
\textbf{Total (Ours)} & 316.4 & 308K \\
\hline
\textbf{Qwen-Audio}\cite{chu2023qwen,chu2024qwen2} & 10000 & - \\
\textbf{SALMONN}\cite{tang2023salmonn} & 1044 & 370K \\
\textbf{LTU-AS}\cite{gong2023joint} & 1784 & 5682K \\
\hline
\end{tabular}
\vspace{-10pt}
\label{tab:dataset_summary}
\end{table}



\begin{table*}[ht]
\footnotesize
\centering
\caption{
Evaluation results of our proposed models and other baseline models.
Acc denotes accuracy, \textbf{F1 (Y)} and \textbf{F1 (N)} are F1 scores for yes and no answers, respectively, and \textbf{F1 (W)} is the weighted F1 score. 
\textbf{Yes} shows the percentage of yes responses. 
Syn-Hyn Test refers to the Synonym and Hypernym Test. 
All values in the table are percentages.
}
\begin{tabular}{l c c c c c c c c c c}
\toprule
\textbf{} & \multicolumn{5}{c}{\textbf{Audio Hallcination}} & \multicolumn{1}{c}{\textbf{ClothoAQA}} & \multicolumn{4}{c}{\textbf{Syn-Hyn Test}} \\
\cmidrule(r){2-6} \cmidrule(r){7-7} \cmidrule(r){8-11}
\textbf{Models} & \textbf{Acc} & \textbf{F1 (Y)} & \textbf{F1 (N)} & \textbf{F1 (W)} & \textbf{Yes} 
& \textbf{Acc} 
& \textbf{Acc} & \textbf{F1 (Y)} & \textbf{F1 (N)} & \textbf{F1 (W)}
\\
\midrule
Qwen-Audio-Chat\cite{chu2023qwen}
& 66.0 & 72.9 & 54.4 & 63.7 & 75.4 
& 74.9 
& 80.0 & \underline{79.8} & 80.3 & 80.1
\\
Qwen2-Audio-Instruct\cite{chu2024qwen2} 
& 67.6 & 75.3 & 53.0 & 64.1 & 81.0 
& 75.3
& 68.7 & 71.7 & 65.0 & 67.7
\\
SALMONN-7B\cite{tang2023salmonn}
& 57.5 & 70.1 & 26.5 & 48.3 & 92.1
& 76.7
& 54.7 & 57.1 & 52.0 & 54.0
\\
SALMONN-13B\cite{tang2023salmonn}
& 70.8 & 77.0 & 60.1 & 68.5 & 76.8
& \underline{84.0}
& 61.1 & 66.9 & 52.8 & 58.6
\\
LTU-AS\cite{gong2023joint} 
& 48.8 & 51.7 & 45.5 & 48.6 & 56.0
& 57.0
& 50.6 & 27.7 & 62.5 & 48.7
\\
Gemini-1.5-Pro\cite{team2024gemini} 
& 68.6 & 65.9 & 71.0 & 68.4 & 42.0
& 68.4
& 78.8 & 72.5 & 82.8 & 78.6
\\
\midrule
\textcolor{gray}{Ours (Gemini, Positive + Positive)} 
& 50.4 & 66.6 & 3.4 & 35.0 & 98.7
& 53.6 
& 48.3 & 56.3 & 36.9 & 44.3
\\
\textcolor{gray}{Ours (Gemini, Positive + Negative)} 
& 52.6 & 63.6 & 32.3 & 47.9 & 80.0
& 61.0
& 69.6 & 29.1 & 80.7 & 62.8
\\
\textcolor{gray}{Ours (Gemini, Combined)} 
& 62.1 & 57.3 & 66.0 & 61.6 & 38.7
& 77.3
& 66.5 & 48.7 & 75.1 & 65.6
\\
\textcolor{gray}{Ours (Rule-based Template)} 
& 54.4 & 70.2 & 2.5 & 36.3 & 99.1
& 56.8
& 37.4 & 49.2 & 18.5 & 28.3
\\
\midrule
Ours (Positive + Positive) 
& 66.3 & 73.1 & 55.2 & 64.1 & 74.8
& 81.5
& \textbf{82.8} & \textbf{81.7} & 83.8 & \textbf{83.0}
\\
Ours (Positive + Negative) 
& \textbf{77.5} & \textbf{80.2} & \textbf{74.0} & \textbf{77.1} & 63.4
& \textbf{84.3}
& \underline{82.2} & \underline{79.8} & \underline{84.1} & \underline{82.2}
\\
Ours (Combined) 
& \underline{74.5} & \underline{77.2} & \underline{71.1} & \underline{74.1} & 61.7
& {83.3}
& 79.7 & 71.0 & \textbf{84.4} & 78.9
\\
\bottomrule
\end{tabular}
\vspace{-8pt}
\label{tab:results}
\end{table*}

\subsection{Evaluation Benchmarks}
\subsubsection{Audio Hallucination}
Previous studies\cite{kuan2024understanding, kuan2024can} have introduced benchmarks to evaluate object hallucination in audio. 
In this task, the model is asked whether a specific sound event is present in the audio. For example, \textit{Is there a car horn sound in the audio?} 
Since the question format is binary, standard classification metrics can be used to assess the model's performance. 
To further analyze the model's behavior, we also calculate precision, recall, and F1 scores separately for questions where the correct answer is ``yes'' and ``no''.

\subsubsection{Audio Question Answering}
Clotho-AQA~\cite{lipping2022clotho} is a widely used dataset for audio question answering (AQA), where systems generate natural language answers from audio signals and related questions. 
Questions are crowd-sourced, and answers come from multiple annotators. 
To ensure accuracy, only questions with unanimous annotator agreement are included.

\subsubsection{Synonym and Hypernym Test}

A previous study~\cite{ccoban2024mllms} introduced a synonym and hypernym test to assess whether models understand audio and its semantic connections to text. 
Synonyms denote similar meanings, while hypernyms represent broader categories encompassing specific terms. These relationships help formulate text prompts for evaluation.
For example, given an audio of a songbird chirping, a possible question could be: \textit{Is the sound from an object that is a type of chordate?} The expected answer is \textit{yes}. 
This benchmark evaluates whether audio-aware LLMs genuinely comprehend audio and reason about semantic relationships or merely rely on keyword associations.
Note that during evaluation, all models use greedy decoding.

\subsection{Ablation Study}
To ensure a fair evaluation of contrastive-like methods, we designed the following experiments:
\begin{enumerate}
    \item \textbf{Positive-only Training Data:} 
    The training set contains only positive samples, with a total of $2N$ data points.
    
    \item \textbf{Positive and Negative Training Data:} 
    The training set includes both positive and negative samples, each with $N$ data points, resulting in a total of $2N$ data points.
    
    \item \textbf{Combined Training Data:} 
    The training set consists of combined samples, as described in Section 3.1. 
    A combined sample includes both sound events that are present and those that are absent within a single sample. 
    In contrast, positive samples only contain present sound events, while negative samples only contain absent sound events. 
    Under the combined setup, there are $N$ data points. 
    Since each data point carries information about both present and absent sound events, it is equivalent to having a total of $2N$ data points.
    
\end{enumerate}

\section{Results}
\subsection{Performance on Audio Hallucination}
In Table~\ref{tab:results}, compared to previous baselines, our proposed models achieve the best performance on the audio hallucination benchmark, both in terms of accuracy and F1 score.
For questions where the ground truth is ``no'', which are designed to test whether the model can accurately identify sound events that do not exist in the given audio, we observed a significant improvement when negative samples were included in our data generation pipeline. 
This suggests that the contrastive-like approach helps mitigate audio hallucination. 
In contrast, improvements from using only positive samples were less noticeable.
Based on our experimental results, separating positive samples and negative samples performs better compared to combining them into a single set of samples.

\subsection{Performance on Audio Question Answering}
While our proposed model performs well on the audio hallucination benchmark, it also delivers competitive results on traditional audio question answering benchmarks, even outperforming other baselines. 
This demonstrates that our approach is not only effective in mitigating hallucinations but also achieves a solid level of performance in basic audio understanding. 


\subsection{Performance on Synonym and Hypernym Test}
One of the key objectives of audio-text alignment is to align audio information, which belongs to a different modality than text, with the text space of large language models. 
Few studies have systematically and specifically explored this issue. 
As an initial exploration, we based our analysis on\cite{ccoban2024mllms} to evaluate whether our model and other baseline models truly understand audio content and use it to reason about semantic relationships.
In Table \ref{tab:results}, our proposed models achieve impressive performance, demonstrating stronger and more robust audio-text alignment compared to other baselines. While Qwen-Audio-Chat achieves comparable performance, we use only 3\% of their dataset's total hours and do not modify any parameters of the audio foundation model or the backbone LLM.

\subsection{Comparison of Data Generation Approaches}
We also examine the scenario of using a stronger LLM to generate synthetic data via our proposed pipeline. 
Specifically, we select Gemini-1.5-Pro to synthesize the data described in Section 3.1. 
We apply the same method but use a more advanced model to generate the data. Experimental results show that using a more capable LLM does not provide additional benefits, indicating that self-generated data from the backbone LLM is already sufficient in this case.
On the other hand, we design several rule-based prompt templates as a comparison method for generating data. 
Using pre-designed description-based prompts, such as \textit{Describe the audio,} the training target corresponds to the ground truth captions of the audio.  
However, experimental results show that this rule-based method performs poorly, conveying the necessity of using self-generated data from the LLM.


\section{Conclusion, Future work, and Limitation}

This paper proposes a contrastive-like training method that enables audio-aware large language models to recognize both present and absent sounds in audio by self-generated data, effectively reducing hallucination issues and improving model reliability. 
Additionally, we achieve impressive results on audio understanding and reasoning benchmarks, demonstrating the robustness and versatility of this approach. 


There are still some limitations to this work.
First, this paper focuses on mitigating audio-related hallucinations, while speech-related hallucinations remain relatively underexplored.
Second, incorporating visual modalities could help reduce hallucinations and enhance audio reasoning tasks in audio-visual scenarios.
We leave these aspects for future work.



\bibliographystyle{IEEEtran}
\bibliography{mybib}

\begin{thebibliography}{10}
\providecommand{\url}[1]{#1}
\csname url@samestyle\endcsname
\providecommand{\newblock}{\relax}
\providecommand{\bibinfo}[2]{#2}
\providecommand{\BIBentrySTDinterwordspacing}{\spaceskip=0pt\relax}
\providecommand{\BIBentryALTinterwordstretchfactor}{4}
\providecommand{\BIBentryALTinterwordspacing}{\spaceskip=\fontdimen2\font plus
\BIBentryALTinterwordstretchfactor\fontdimen3\font minus \fontdimen4\font\relax}
\providecommand{\BIBforeignlanguage}[2]{{%
\expandafter\ifx\csname l@#1\endcsname\relax
\typeout{** WARNING: IEEEtran.bst: No hyphenation pattern has been}%
\typeout{** loaded for the language `#1'. Using the pattern for}%
\typeout{** the default language instead.}%
\else
\language=\csname l@#1\endcsname
\fi
#2}}
\providecommand{\BIBdecl}{\relax}
\BIBdecl

\bibitem{kuan2024can}
C.-Y. Kuan and H.-y. Lee, ``Can large audio-language models truly hear? tackling hallucinations with multi-task assessment and stepwise audio reasoning,'' in \emph{ICASSP 2025-2025 IEEE International Conference on Acoustics, Speech and Signal Processing (ICASSP)}.\hskip 1em plus 0.5em minus 0.4em\relax IEEE, 2025.

\bibitem{kuan2024understanding}
C.-Y. Kuan, W.-P. Huang, and H.-y. Lee, ``Understanding sounds, missing the questions: The challenge of object hallucination in large audio-language models,'' \emph{Interspeech 2024}, 2024.

\bibitem{rawte2023survey}
V.~Rawte, A.~Sheth, and A.~Das, ``A survey of hallucination in large foundation models,'' \emph{arXiv:2309.05922}, 2023.

\bibitem{zhang2023siren}
Y.~Zhang \emph{et~al.}, ``Siren's song in the ai ocean: a survey on hallucination in large language models,'' \emph{arXiv:2309.01219}, 2023.

\bibitem{huang2023survey}
L.~Huang \emph{et~al.}, ``A survey on hallucination in large language models: Principles, taxonomy, challenges, and open questions,'' \emph{arXiv:2311.05232}, 2023.

\bibitem{friel2023chainpoll}
R.~Friel and A.~Sanyal, ``Chainpoll: A high efficacy method for llm hallucination detection,'' \emph{arXiv preprint arXiv:2310.18344}, 2023.

\bibitem{liu2024survey}
H.~Liu, W.~Xue, Y.~Chen, D.~Chen, X.~Zhao, K.~Wang, L.~Hou, R.~Li, and W.~Peng, ``A survey on hallucination in large vision-language models,'' \emph{arXiv preprint arXiv:2402.00253}, 2024.

\bibitem{fathullah2023towards}
Y.~Fathullah \emph{et~al.}, ``Towards general-purpose speech abilities for large language models using unpaired data,'' \emph{arXiv preprint arXiv:2311.06753}, 2023.

\bibitem{lu2024developing}
K.-H. Lu \emph{et~al.}, ``Developing instruction-following speech language model without speech instruction-tuning data,'' \emph{arXiv preprint arXiv:2409.20007}, 2024.

\bibitem{wang2023blsp}
C.~Wang \emph{et~al.}, ``Blsp: Bootstrapping language-speech pre-training via behavior alignment of continuation writing,'' \emph{arXiv preprint arXiv:2309.00916}, 2023.

\bibitem{wang2024blsp}
------, ``Blsp-emo: Towards empathetic large speech-language models,'' \emph{arXiv preprint arXiv:2406.03872}, 2024.

\bibitem{chu2023qwen}
Y.~Chu, J.~Xu, X.~Zhou, Q.~Yang, S.~Zhang, Z.~Yan, C.~Zhou, and J.~Zhou, ``Qwen-audio: Advancing universal audio understanding via unified large-scale audio-language models,'' \emph{arXiv preprint arXiv:2311.07919}, 2023.

\bibitem{chu2024qwen2}
Y.~Chu, J.~Xu, Q.~Yang, H.~Wei, X.~Wei, Z.~Guo, Y.~Leng, Y.~Lv, J.~He, J.~Lin \emph{et~al.}, ``Qwen2-audio technical report,'' \emph{arXiv preprint arXiv:2407.10759}, 2024.

\bibitem{tang2023salmonn}
C.~Tang, W.~Yu, G.~Sun, X.~Chen, T.~Tan, W.~Li, L.~Lu, Z.~Ma, and C.~Zhang, ``Salmonn: Towards generic hearing abilities for large language models,'' \emph{arXiv preprint arXiv:2310.13289}, 2023.

\bibitem{gong2023joint}
Y.~Gong, A.~H. Liu, H.~Luo, L.~Karlinsky, and J.~Glass, ``Joint audio and speech understanding,'' in \emph{2023 IEEE Automatic Speech Recognition and Understanding Workshop (ASRU)}.\hskip 1em plus 0.5em minus 0.4em\relax IEEE, 2023, pp. 1--8.

\bibitem{hulora}
E.~J. Hu, P.~Wallis, Z.~Allen-Zhu, Y.~Li, S.~Wang, L.~Wang, W.~Chen \emph{et~al.}, ``Lora: Low-rank adaptation of large language models,'' in \emph{International Conference on Learning Representations}, 2021.

\bibitem{zhuminigpt}
D.~Zhu, J.~Chen, X.~Shen, X.~Li, and M.~Elhoseiny, ``Minigpt-4: Enhancing vision-language understanding with advanced large language models,'' in \emph{The Twelfth International Conference on Learning Representations}, 2023.

\bibitem{lu2024desta}
K.-H. Lu \emph{et~al.}, ``Desta: Enhancing speech language models through descriptive speech-text alignment,'' \emph{arXiv preprint arXiv:2406.18871}, 2024.

\bibitem{kuan2024speech}
C.-Y. Kuan, C.-K. Yang, W.-P. Huang, K.-H. Lu, and H.-y. Lee, ``Speech-copilot: Leveraging large language models for speech processing via task decomposition, modularization, and program generation,'' \emph{arXiv preprint arXiv:2407.09886}, 2024.

\bibitem{huang2024dynamic}
C.-y. Huang \emph{et~al.}, ``Dynamic-superb: Towards a dynamic, collaborative, and comprehensive instruction-tuning benchmark for speech,'' in \emph{ICASSP 2024-2024 IEEE International Conference on Acoustics, Speech and Signal Processing (ICASSP)}.\hskip 1em plus 0.5em minus 0.4em\relax IEEE, 2024, pp. 12\,136--12\,140.

\bibitem{huang2024dynamic2}
C.-y. Huang, W.-C. Chen \emph{et~al.}, ``Dynamic-superb phase-2: A collaboratively expanding benchmark for measuring the capabilities of spoken language models with 180 tasks,'' \emph{arXiv preprint arXiv:2411.05361}, 2024.

\bibitem{achiam2023gpt}
J.~Achiam, S.~Adler, S.~Agarwal, L.~Ahmad, I.~Akkaya, F.~L. Aleman, D.~Almeida, J.~Altenschmidt, S.~Altman, S.~Anadkat \emph{et~al.}, ``Gpt-4 technical report,'' \emph{arXiv preprint arXiv:2303.08774}, 2023.

\bibitem{radford2023robust}
A.~Radford \emph{et~al.}, ``Robust speech recognition via large-scale weak supervision,'' in \emph{International conference on machine learning}.\hskip 1em plus 0.5em minus 0.4em\relax PMLR, 2023, pp. 28\,492--28\,518.

\bibitem{gong_whisperat}
Y.~Gong, S.~Khurana, L.~Karlinsky, and J.~Glass, ``Whisper-at: Noise-robust automatic speech recognizers are also strong audio event taggers,'' in \emph{Proc. Interspeech 2023}, 2023.

\bibitem{ma2023investigating}
R.~Ma, A.~Liusie, M.~J. Gales, and K.~M. Knill, ``Investigating the emergent audio classification ability of asr foundation models,'' \emph{arXiv preprint arXiv:2311.09363}, 2023.

\bibitem{dubey2024llama}
A.~Dubey \emph{et~al.}, ``The llama 3 herd of models,'' \emph{arXiv preprint arXiv:2407.21783}, 2024.

\bibitem{li2023blip}
J.~Li, D.~Li, S.~Savarese, and S.~Hoi, ``Blip-2: Bootstrapping language-image pre-training with frozen image encoders and large language models,'' in \emph{International conference on machine learning}.\hskip 1em plus 0.5em minus 0.4em\relax PMLR, 2023, pp. 19\,730--19\,742.

\bibitem{7952261}
J.~F. Gemmeke, D.~P.~W. Ellis, D.~Freedman, A.~Jansen, W.~Lawrence, R.~C. Moore, M.~Plakal, and M.~Ritter, ``Audio set: An ontology and human-labeled dataset for audio events,'' in \emph{2017 IEEE International Conference on Acoustics, Speech and Signal Processing (ICASSP)}, 2017, pp. 776--780.

\bibitem{kim2019audiocaps}
C.~D. Kim, B.~Kim, H.~Lee, and G.~Kim, ``Audiocaps: Generating captions for audios in the wild,'' in \emph{Proceedings of the 2019 Conference of the North American Chapter of the Association for Computational Linguistics: Human Language Technologies, Volume 1 (Long and Short Papers)}, 2019, pp. 119--132.

\bibitem{fonseca2021fsd50k}
E.~Fonseca, X.~Favory, J.~Pons, F.~Font, and X.~Serra, ``Fsd50k: an open dataset of human-labeled sound events,'' \emph{IEEE/ACM Transactions on Audio, Speech, and Language Processing}, vol.~30, pp. 829--852, 2021.

\bibitem{8682858}
I.~Martín-Morató, A.~Mesaros, T.~Heittola, T.~Virtanen, M.~Cobos, and F.~J. Ferri, ``Sound event envelope estimation in polyphonic mixtures,'' in \emph{ICASSP 2019 - 2019 IEEE International Conference on Acoustics, Speech and Signal Processing (ICASSP)}, 2019, pp. 935--939.

\bibitem{piczak2015dataset}
\BIBentryALTinterwordspacing
K.~J. Piczak, ``{ESC}: {Dataset} for {Environmental Sound Classification},'' in \emph{Proceedings of the 23rd {Annual ACM Conference} on {Multimedia}}.\hskip 1em plus 0.5em minus 0.4em\relax {ACM Press}, 2015, pp. 1015--1018. [Online]. Available: \url{http://dl.acm.org/citation.cfm?doid=2733373.2806390}
\BIBentrySTDinterwordspacing

\bibitem{Salamon:UrbanSound:ACMMM:14}
J.~Salamon, C.~Jacoby, and J.~P. Bello, ``A dataset and taxonomy for urban sound research,'' in \emph{22nd {ACM} International Conference on Multimedia (ACM-MM'14)}, Orlando, FL, USA, Nov. 2014, pp. 1041--1044.

\bibitem{lipping2022clotho}
S.~Lipping, P.~Sudarsanam, K.~Drossos, and T.~Virtanen, ``Clotho-aqa: A crowdsourced dataset for audio question answering,'' in \emph{2022 30th European Signal Processing Conference (EUSIPCO)}.\hskip 1em plus 0.5em minus 0.4em\relax IEEE, 2022, pp. 1140--1144.

\bibitem{gong_vocalsound}
Y.~Gong, J.~Yu, and J.~Glass, ``Vocalsound: A dataset for improving human vocal sounds recognition,'' in \emph{ICASSP 2022 - 2022 IEEE International Conference on Acoustics, Speech and Signal Processing (ICASSP)}, 2022, pp. 151--155.

\bibitem{team2024gemini}
G.~Team, P.~Georgiev, V.~I. Lei, R.~Burnell, L.~Bai, A.~Gulati, G.~Tanzer, D.~Vincent, Z.~Pan, S.~Wang \emph{et~al.}, ``Gemini 1.5: Unlocking multimodal understanding across millions of tokens of context,'' \emph{arXiv preprint arXiv:2403.05530}, 2024.

\bibitem{ccoban2024mllms}
E.~B. {\c{C}}oban, M.~I. Mandel, and J.~Devaney, ``What do mllms hear? examining the interaction between llm and audio encoder components in multimodal large language models,'' in \emph{Audio Imagination: NeurIPS 2024 Workshop AI-Driven Speech, Music, and Sound Generation}, 2024.

\end{thebibliography}

\end{document}